# Solution-Based Synthesis of Fe–Co–Ni Prussian Blue Analogue Powders: A Comparative Structural, Spectroscopic and Thermal Study

Larissa de O. Garcia[a,*], Lindiomar Borges de Avila Jr[b], Mengmeng Wang[c], Yann Garcia[c], Christian K. Müller[a] and Flavio Abreu Araujo[b,*]

[a]*Faculty of Physical Engineering/Computer Sciences, University of Applied Sciences Zwickau, Zwickau, 08056, Germany*
[b]*Institute of Condensed Matter and Nanosciences (IMCN), Université catholique de Louvain (UCLouvain), Louvain-la-Neuve, B-1348, Belgium*
[c]*Institute of Condensed Matter and Nanosciences, Molecular Chemistry, Materials and Catalysis (IMCN/MOST), Université catholique de Louvain (UCLouvain), Louvain-la-Neuve, B-1348, Belgium*

ABSTRACT

Prussian Blue (Fe-PBA) and its Co and Ni analogues (Co-PBA, Ni-PBA) were prepared by the same additive-free aqueous co-precipitation of $K_4[Fe(CN)_6]$ with $FeCl_3$, $CoCl_2$ or $NiCl_2$, and compared by X-ray diffraction, ATR-FTIR and Raman spectroscopy, FEG-SEM/EDS and TG/SDTA. All three powders adopt the cubic cyanide-bridged PBA framework, with lattice parameters of 10.10 ± 0.02, 10.01 ± 0.02 and 10.10 ± 0.03 Å estimated from three reflections. The dominant infrared C≡N band moves from 2062 to 2071 and 2087 $cm^{-1}$ along the series. Fe-PBA, obtained from an Fe(III) salt, contains less potassium than the analogues obtained from Co(II) and Ni(II) salts and loses more mass below 200 °C. Sharp KCl reflections in all three diffraction patterns, and further unassigned reflections for Ni-PBA, show crystalline by-products that washing by decantation did not remove. The oxidation state of the metal precursor and the washing step therefore have to be considered together with the identity of the transition metal when the three compositions are compared.

## 1. Introduction

Prussian Blue analogues (PBAs) are cyanide-bridged coordination materials characterized by open three-dimensional frameworks and substantial compositional flexibility [1–3]. They can be represented by the generalized composition $A_x M[M'(CN)_6]_{1-y}\square_y \cdot nH_2O$, where $M$ and $M'$ are transition-metal centers, $A^+$ represents interstitial cations, $\square$ denotes vacancies in the $M'(CN)_6$ sublattice, and $n$ is the number of water molecules per formula unit. Variations in transition-metal identity, vacancy concentration, cation content, and hydration can modify the local coordination environment and the average crystal structure [4, 5]. These characteristics underpin the electrochemical, magnetic, catalytic, gas-sorption and ion-transport properties reported for PBAs [3–9], and related cyanide-bridged frameworks show spin-state switching [10, 11]. These structural degrees of freedom influence local coordination, lattice dimensions, defect chemistry, hydration, and charge-transport processes, supporting the use of PBAs in electrochemical energy storage, sensing, catalysis, and electronic applications [3, 12–14]. PBA-based materials have also attracted interest for resistive-switching and memristive devices, where ionic redistribution and redox processes can contribute to the electrical response [15–23].

Transition-metal substitution provides a direct means of modifying these structural and physicochemical characteristics [24]. Fe-, Co-, and Ni-containing hexacyanoferrates are particularly useful for comparative studies because they can retain the same fundamental cyanide-bridged framework while exhibiting differences in local bonding, lattice dimensions, morphology, hydration, and structural order. Previous work on electrodeposited FeHCF, CoHCF, and NiHCF thin films demonstrated composition-dependent differences in morphology, elemental composition, lattice parameters, and cyanide bonding despite preservation of the cubic framework [25–27]. Such composition-dependent differences may also influence the hydration and thermal response of the corresponding powders.

*Corresponding author
larissa.de.oliveira.garcia@whz.de (L. de O. Garcia); flavio.abreuaraujo@uclouvain.be (F. Abreu Araujo)
ORCID(s): 0000-0003-0267-6488 (L. de O. Garcia)

From a processing perspective, the complexity of a synthesis route is itself an important consideration. Control of PBA nucleation and particle growth can be improved by introducing complexing or chelating agents, such as citrate, or other surfactants and capping species, which can modify crystallization kinetics and particle morphology [28–30]. However, these additives introduce additional chemical variables and may complicate subsequent washing and purification. A deliberately additive-free precipitation route therefore provides a useful complementary strategy when the objective is to obtain structurally identifiable PBA materials using a simple and reproducible protocol. A consequence of this approach is that particle nucleation, growth, and aggregation are less strongly controlled, and the resulting powders may exhibit broader morphological distributions [28, 30]. This is a trade-off between morphological control and synthetic simplicity.

The practical relevance of such a low-complexity route also depends on whether the resulting material retains the characteristic PBA framework. The relationship between synthesis conditions, framework structure, defects, and morphology has been widely documented for PBAs [3, 28, 30]. If the cyanide-bridged structure is preserved despite particle aggregation, the powders provide a useful basis for comparing composition-dependent properties independently of the morphological limitations introduced by processing. Morphology and framework formation should therefore be treated as distinct aspects of the material: deposit quality depends on processing conditions, whereas framework preservation is assessed from complementary structural and spectroscopic measurements.

A reliable assessment of these relationships requires characterization across different aspects and length scales of the material. FEG-SEM/EDS provides information on particle morphology and elemental composition, while ATR-FTIR and Raman spectroscopy probe the local vibrational environment of the cyanide framework and associated water species [31, 32]. X-ray diffraction (XRD) probes the average long-range structure and provides evidence for the retention of the characteristic PBA crystal framework [28, 33]. Thermogravimetry combined with single differential thermal analysis (TG/SDTA) follows mass changes and thermal events as a function of temperature [30, 34]. It is particularly important to combine these techniques because PBAs can contain different populations of water, including surface-adsorbed, interstitial, and coordination-associated water, which may be released over different temperature ranges [28, 33]. Consequently, TG mass loss alone does not necessarily provide an unambiguous assignment of individual water populations or distinguish dehydration from irreversible framework degradation. Complementary post-treatment structural and spectroscopic characterization is therefore important for determining whether mass-loss events are associated with reversible dehydration or permanent structural changes [30, 33, 34].

Despite the extensive literature on PBAs, comparative studies of Fe-, Co-, and Ni-containing powders prepared under identical conditions remain useful for isolating composition-dependent effects. Keeping pH, aging, washing, and drying conditions consistent reduces sources of variation unrelated to transition-metal identity and provides a controlled basis for comparing morphology, local structure, hydration, and thermal response. One variable cannot be held constant in such a series: here Prussian Blue is made from an Fe(III) salt and the Co and Ni analogues from M(II) salts, so at equal vacancy content the Co and Ni frameworks need one more $K^+$ per formula unit, and potassium uptake and vacancy content differ by construction [2, 28].

In this work, Fe-, Co-, and Ni-based Prussian Blue analogue powders were synthesized by a straightforward aqueous co-precipitation route under identical conditions, without citrate or other nucleation-controlling additives. The three compositions were systematically compared by FEG-SEM/EDS, ATR-FTIR, Raman spectroscopy, X-ray diffraction (XRD), and TG/SDTA to establish how transition-metal identity influences chemical composition, cyanide coordination, hydration, crystal structure, morphology, and thermal behavior. Particular attention was given to distinguishing composition-dependent features from morphological changes introduced during powder processing and drop casting.

## 2. Materials and Methods

### 2.1. Synthesis of Prussian Blue Analogue Powders

Prussian Blue analogue (PBA) powders containing Fe, Co, and Ni as the transition-metal components were synthesized by aqueous co-precipitation. Potassium ferrocyanide, $K_4[Fe(CN)_6]$, was used as the cyanometallate precursor, while iron(III), cobalt(II), and nickel(II) chlorides were used as the corresponding transition-metal precursors, namely $FeCl_3 \cdot 6H_2O$, $CoCl_2 \cdot 6H_2O$, and $NiCl_2 \cdot 6H_2O$, respectively. Iron is introduced as Fe(III), whereas Co and Ni enter as M(II). Fe-PBA is therefore Prussian Blue, with Fe(III)–N≡C–Fe(II) linkages, and its framework charge is balanced differently from that of the Co and Ni analogues. For each composition, 25 mmol of potassium ferrocyanide and 25 mmol of the corresponding transition-metal chloride were used. This 1:1 precursor feed ratio was kept identical across the series as an experimental synthesis condition and is not taken to define the stoichiometry of the isolated solid. A 1:1 ratio corresponds to the potassium-rich limits $KFe^{III}[Fe^{II}(CN)_6]$ and $K_2M^{II}[Fe^{II}(CN)_6]$ (M = Co, Ni); the potassium-free limits, $Fe^{III}_4[Fe^{II}(CN)_6]_3$ and $M^{II}_2[Fe^{II}(CN)_6]$, contain more metal than hexacyanoferrate, and the isolated solids can lie anywhere between these limits or be partly oxidized, as in $KCo^{III}[Fe^{II}(CN)_6]$. The actual metal/cyanometallate/vacancy/counter-ion composition was not independently determined by a bulk quantitative method in the present study. All reagents were purchased from Sigma-Aldrich and used as received.

The precursor solutions were prepared separately in deionized water, and the pH was adjusted to approximately 2 using hydrochloric acid. Because acidification of ferrocyanide-containing solutions can generate hazardous cyanide-containing gases under sufficiently acidic conditions, pH adjustment and subsequent heating were performed under appropriate laboratory safety controls and ventilation. The transition-metal chloride solution was added dropwise to the potassium ferrocyanide solution under continuous magnetic stirring at room temperature. After complete addition, the resulting suspension was aged at 70 °C for 2 h. No citrate, complexing or chelating agents, surfactants, or capping additives were used. Following aging, the suspension was allowed to settle, and the supernatant was removed by decantation. The precipitates were washed several times with deionized water by repeated resuspension, settling, and decantation to remove soluble by-products and residual precursor species. No centrifugation or filtration was used, and because decantation leaves part of the supernatant in the wet sediment, dissolved by-products such as KCl can crystallize in the powders during drying. The washed precipitates were dried at 70 °C overnight. The resulting powders were denoted as Fe-PBA, Co-PBA, and Ni-PBA according to the transition-metal component.

## 2.2. Characterization

The synthesized powders were characterized using X-ray diffraction (XRD), Fourier-transform infrared spectroscopy (FTIR), Raman spectroscopy, field-emission scanning electron microscopy (FEG-SEM), energy-dispersive X-ray spectroscopy (EDS), and thermogravimetry with single differential thermal analysis (TG/SDTA) to evaluate their average crystal structure, local chemical environment, morphology, elemental composition, hydration-related mass loss, and thermal behavior.

X-ray diffraction measurements were performed using a PANalytical X'Pert MRD diffractometer with Co K$\alpha$ radiation ($\lambda$ = 1.78896 Å) in Bragg–Brentano geometry. Diffraction patterns were collected over a 2$\theta$ range of 10°–80°. The diffraction data were used to identify the crystalline PBA framework and evaluate differences in the structural characteristics of the Fe-, Co-, and Ni-containing powders.

FTIR spectra were acquired using a Shimadzu IRSpirit-T spectrometer equipped with an attenuated total reflection (ATR) accessory. Measurements were performed directly on the synthesized powders.

Surface morphology and microstructure were examined by field-emission scanning electron microscopy (FEG-SEM, TESCAN CLARA, Brno, Czech Republic). For SEM, EDS, and Raman measurements, the powders were dispersed in an 80:20 vol% water/ethanol mixture and deposited by drop casting onto Au (50 nm)/Cr (5 nm)-coated Si substrates. This preparation was used to immobilize the powder particles and provide a more stable and reproducible surface for microscopic and spectroscopic measurements. No binder, surfactant, or additional film-forming additive was used. The deposits were allowed to dry under ambient conditions before characterization. Since solvent evaporation during drop casting may induce particle rearrangement, aggregation, shrinkage, or cracking, the resulting SEM images were interpreted primarily in terms of qualitative particle morphology and deposit structure rather than as a direct representation of the intrinsic powder packing or particle-size distribution.

Elemental composition was determined by energy-dispersive X-ray spectroscopy (EDS, Oxford Instruments) integrated with the FEG-SEM and operated at an accelerating voltage of 20 kV. Measurements were performed on the drop-cast powder deposits. Given the semi-quantitative nature of EDS and the presence of the underlying Au/Cr-coated Si substrate, the measured elemental contents were interpreted primarily in terms of relative composition and spatial distribution rather than absolute stoichiometry. Contributions from the substrate and the finite electron-beam interaction volume were considered when interpreting the measured elemental contents.

Raman spectra were acquired using a confocal Raman microscope (WITec RISE, Ulm, Germany) equipped with a 532 nm excitation laser and a Zeiss LD EC Epiplan-Neofluar 100×/0.75 objective. Measurements were performed on the same drop-cast powder deposits used for SEM and EDS characterization. The laser power was limited to 0.4 mW to minimize local heating and reduce the possibility of laser-induced modification of the samples. This is above the 0.005–0.06 mW range that Moretti and Gervais found free of laser-induced changes for Prussian Blue [35]. Spectra were collected using a 600 grooves mm$^{-1}$ grating, an integration time of 0.7 s, and 100 accumulations. All Raman measurements were performed under identical acquisition conditions.

Thermal analysis was performed using a Mettler Toledo TGA/SDTA 851e analyzer under flowing $N_2$ at a gas flow rate of 100 mL $min^{-1}$. Approximately 12.48, 9.75, and 6.91 mg of the Fe-, Co-, and Ni-containing powders, respectively, were analyzed. The samples were heated from 25 to 600 °C at a constant heating rate of 10 °C $min^{-1}$. The thermogravimetric data were evaluated as a function of the measured sample temperature ($T_s$), with emphasis on the temperature range and magnitude of the observed mass-loss events, particularly those associated with dehydration and subsequent thermal evolution of the PBA framework. The simultaneously recorded SDTA signal, the difference between the measured sample temperature and the program temperature, was used to locate thermal events.

# 3. Results and Discussion

## 3.1. Morphology and Elemental Composition

Representative FEG-SEM micrographs of the Fe-, Co-, and Ni-based PBA powders are shown in Fig. 1. All three compositions exhibit pronounced aggregation, with particulate features forming interconnected secondary agglomerates. Qualitative differences in aggregate texture are visible among the samples, but primary particles are not resolved at these magnifications, and the images describe the dried deposits, not the primary-particle size distribution.

Because the powders were dispersed in an 80:20 vol% water/ethanol mixture and drop cast onto Au (50 nm)/Cr (5 nm)-coated Si substrates before analysis, the observed morphology represents the deposited material rather than isolated primary particles. Solvent evaporation can promote particle redistribution, aggregation, and rearrangement within the drying droplet [36–39].

Interconnected fissures are also visible across the deposited regions. These features are attributed primarily to drying-induced shrinkage and residual stresses within the particulate layer rather than fracture of individual PBA particles [40–42]. Apart from these cracks, the deposits are continuous at the scale of Fig. 1, although no binder, surfactant or other film-forming additive was used. Their structure depends on solvent composition, particle loading, drying conditions and substrate interactions [40, 41]. thickness and porosity were not measured.

The corresponding EDS elemental maps are presented in Fig. 2. Carbon and nitrogen are distributed throughout the analyzed regions, consistent with the cyanide-bridged framework. Fe is detected in all three compositions, together with Co or Ni in the corresponding samples, as expected for the Co- and Ni-containing hexacyanoferrate frameworks [3, 6]. The cracks appear dark in every map, so the contrast in Fig. 2 is dominated by topography. Apart from the cracks and the brighter K and Cl regions in Ni-PBA, no other element shows pronounced segregation at this scale.

The semi-quantitative EDS compositions are summarized in Table 1. The values should not be interpreted as absolute stoichiometric compositions because EDS quantification is affected by light-element detection, particle morphology and geometry, substrate contributions, and the effective X-ray interaction volume [43, 44].Therefore, the data are primarily used to identify the constituent elements and compare their relative distributions among the three compositions. The measured Co/Fe and Ni/Fe atomic ratios are 1.06 and 0.59. In the general formula $A_xM[Fe(CN)_6]_{1-y}\square_y$, $M/Fe = 1/(1-y)$ cannot fall below 1, because vacancies occur only on the hexacyanoferrate sites. The Co/Fe ratio is therefore compatible with a nearly vacancy-free framework, whereas the Ni/Fe ratio is incompatible with a single nickel hexacyanoferrate phase and points to a quantification bias, an additional Fe-containing phase (Section 3.4), or both.

Potassium is detected in all samples, with concentrations of 1.44, 5.54, and 4.87 at.% for Fe-PBA, Co-PBA, and Ni-PBA, respectively. Its presence may reflect potassium-containing species associated with the PBA framework and/or residual potassium-containing material from synthesis and washing. The contents follow the precursor chemistry: at equal vacancy content, an M(II)–N≡C–Fe(II) framework needs one more $K^+$ per formula unit than Prussian Blue, whose potassium-free form, $Fe^{III}_4[Fe^{II}(CN)_6]_3$, balances its charge with vacancies. In Co- and Ni-PBA, K exceeds Cl by 5.1 and 4.2 at.%, so most of the potassium cannot be present as KCl, in Fe-PBA, all of it could be.

Chlorine occurs at lower concentrations of 0.47–1.67 at.%. The use of chloride-containing metal precursors and hydrochloric acid during synthesis provides plausible sources of residual chloride-containing species. Since EDS cannot identify the chemical form of chlorine, the Cl signal is considered a processing-related contribution rather than evidence of chloride incorporation into the cyanide-bridged framework. The accompanying XRD evidence is discussed separately below.

The O concentrations are 10.23, 3.14, and 5.38 at.% for Fe-PBA, Co-PBA, and Ni-PBA, respectively. Oxygen may originate from molecular or surface-associated water and other oxygen-containing species, which cannot be distinguished by EDS alone. Therefore, oxygen content is not used as a quantitative measure of hydration. Hydration is evaluated independently from the TG/SDTA data.

The EDS results revealed the expected framework elements, with no significant macroscopic segregation observed. The K and Cl contributions are consistent with the presence of residual potassium- and chloride-containing material. Along with the additional XRD reflections discussed below, the data are consistent with the presence of residual KCl. In Co-PBA, however, only a small amount of potassium can be attributed to KCl. In Ni-PBA, the rest may be in the additional phase (Section 3.4). The variation in oxygen content cannot be uniquely assigned to differences in hydration. Together with the spectroscopic and diffraction data, these results support the formation of cyanide-bridged PBA-type frameworks, though they do not establish a quantitative bulk formula.

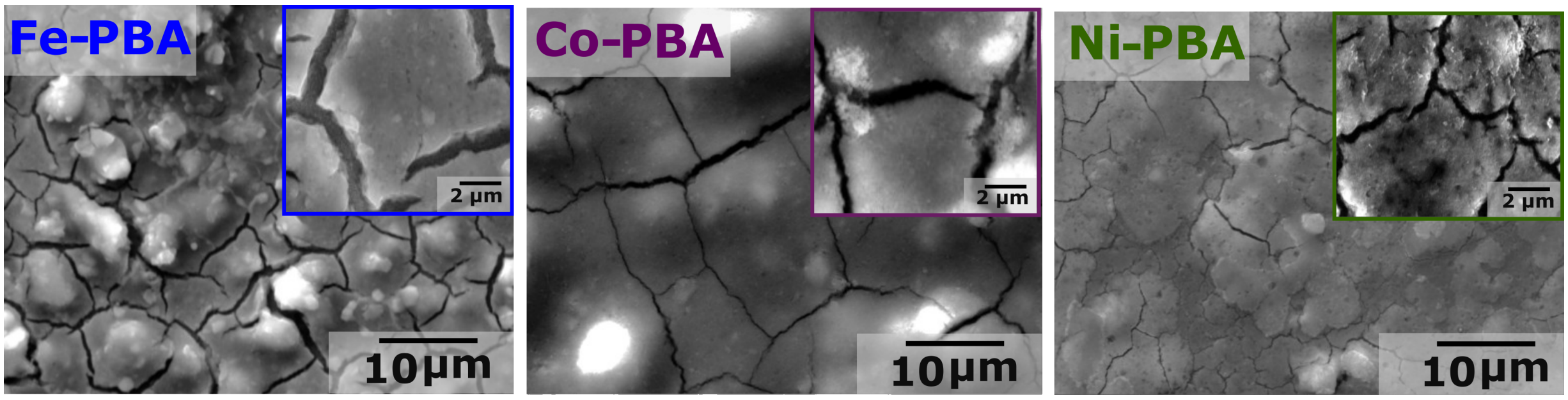

**Figure 1:** Representative FEG-SEM micrographs of Fe-, Co-, and Ni-based PBA powders deposited by drop casting from an 80:20 vol% water/ethanol dispersion. Pronounced aggregation and interconnected drying-induced fissures are observed.

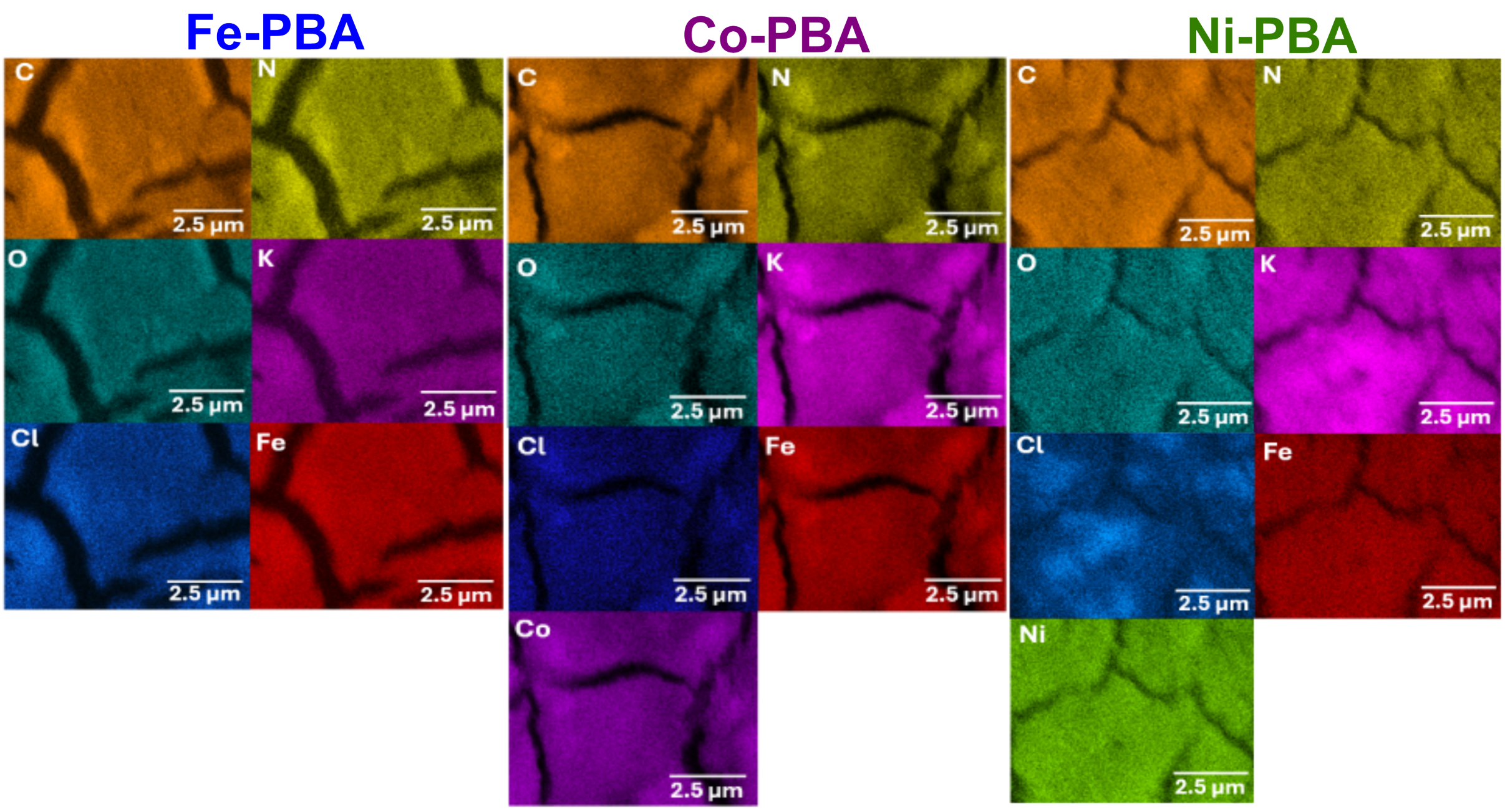

**Figure 2:** Representative EDS elemental maps of Fe-, Co-, and Ni-based PBA deposits showing the spatial distribution of C, N, O, K, Cl, and the corresponding transition-metal component.

**Table 1**
Atomic concentrations obtained from EDS analysis of Fe-, Co-, and Ni-based PBA deposits

| | Atomic concentration (at.%) | | |
|---|---|---|---|
| **Element** | **Fe-PBA** | **Co-PBA** | **Ni-PBA** |
| *Framework and light elements* | | | |
| C | 49.98 | 44.29 | 47.52 |
| N | 29.85 | 35.89 | 36.80 |
| O | 10.23 | 3.14 | 5.38 |
| *Interstitial / residual species* | | | |
| K | 1.44 | 5.54 | 4.87 |
| Cl | 1.67 | 0.47 | 0.66 |
| *Transition metals* | | | |
| Fe | 6.84 | 5.19 | 3.00 |
| Co | | 5.48 | |
| Ni | | | 1.78 |

Empty entries indicate that the corresponding element was not detected within the EDS analysis.

## 3.2. FTIR Spectroscopic Characterization

The ATR-FTIR spectra of the synthesized Fe-, Co-, and Ni-based Prussian Blue analogue (PBA) powders are presented in Figure 3, and the principal absorption bands are summarized in Table 2. All three compositions exhibit the characteristic vibrational features of cyanide-bridged PBA frameworks, including O–H stretching, H–O–H bending, C≡N stretching, and lower-wavenumber metal–cyanide vibrations [31, 32, 45]. The common spectral features confirm preservation of the fundamental PBA framework, while differences in band position and relative intensity indicate composition-dependent changes in the local chemical environment [31, 46].

The high-wavenumber region contains a broad O–H stretching absorption, with higher-wavenumber components at 3634, 3630, and 3646 $cm^{-1}$ for Fe-, Co-, and Ni-PBA, respectively. Corresponding H–O–H bending bands occur at 1599, 1604, and 1608 $cm^{-1}$. These features are consistent with water associated with the PBA powders [32, 47]. In PBA structures, water can occur as interstitial or zeolitic water, coordination-associated water at vacancy-related metal sites, and surface-adsorbed species [47, 48]. The broad O–H absorption therefore cannot be assigned to a single water population from FTIR alone. The spectra nevertheless establish the presence of water-associated species, while their temperature-dependent release is examined independently by TG/SDTA.

The most prominent composition-dependent feature occurs in the C≡N stretching region. The dominant absorption bands are located at 2062, 2071, and 2087 $cm^{-1}$ for Fe-, Co-, and Ni-PBA, respectively, corresponding to a progressive increase in frequency across the series. Since the C≡N stretching frequency is sensitive to the electronic and coordination environment of the cyanide ligand, this shift indicates that transition-metal substitution modifies the local metal–cyanide environment [31, 32, 45]. Ni-PBA shows two further resolved bands, at 2169 and 2055 $cm^{-1}$ (Fig. 3, inset).

The C≡N region may contain contributions from more than one local coordination environment. Structural vacancies and mixed-valence metal centers can generate chemically distinct metal–cyanide configurations within the same PBA framework. For example, NiFe-PBAs exhibit C≡N bands near 2170 and 2098 $cm^{-1}$, which have been assigned to Ni(II)–N≡C–Fe(III) and Ni(II)–N≡C–Fe(II) configurations, respectively [49, 50]. In the present samples, the dominant C≡N bands are well defined across the series. Notably, the additional band observed at 2169 $cm^{-1}$ for Ni-PBA lies close to the characteristic frequency reported for Ni(II)–N≡C–Fe(III), suggesting the possible presence of oxidized ferrocyanide species. To further investigate this possibility, XPS spectroscopic analysis was performed. However, the observed frequency evolution is interpreted primarily as a composition-dependent modification of the cyanide environment rather than as direct evidence of a specific oxidation-state distribution.

It should be noted that FTIR spectroscopy alone cannot unambiguously distinguish the individual contributions of metal identity, oxidation state, vacancy concentration, hydration, and local lattice distortion to the C≡N stretching frequency. Therefore, the spectral changes observed in this region are considered indicative of modifications in the local coordination environment, while oxidation-state assignments require complementary characterization.

A weak feature near 1420 $cm^{-1}$ is also visible in the Fe-PBA spectrum. Since the present dataset does not provide sufficient evidence for a unique vibrational or impurity assignment, this feature is not assigned further.

Additional composition-dependent features are observed in the lower-wavenumber region. Fe-PBA exhibits bands near 603 and 489 $cm^{-1}$, whereas Co-PBA shows features near 591, 501, and 460 $cm^{-1}$. Ni-PBA displays bands near 594 and 432 $cm^{-1}$. These features are associated with vibrations of the metal–cyanide framework and reflect changes in the local lattice dynamics induced by transition-metal substitution [32]. Because the bands are relatively weak and may contain overlapping contributions, individual peaks are not assigned to specific normal modes. Instead, their systematic variation across the series is considered supporting evidence for changes in the local framework environment.

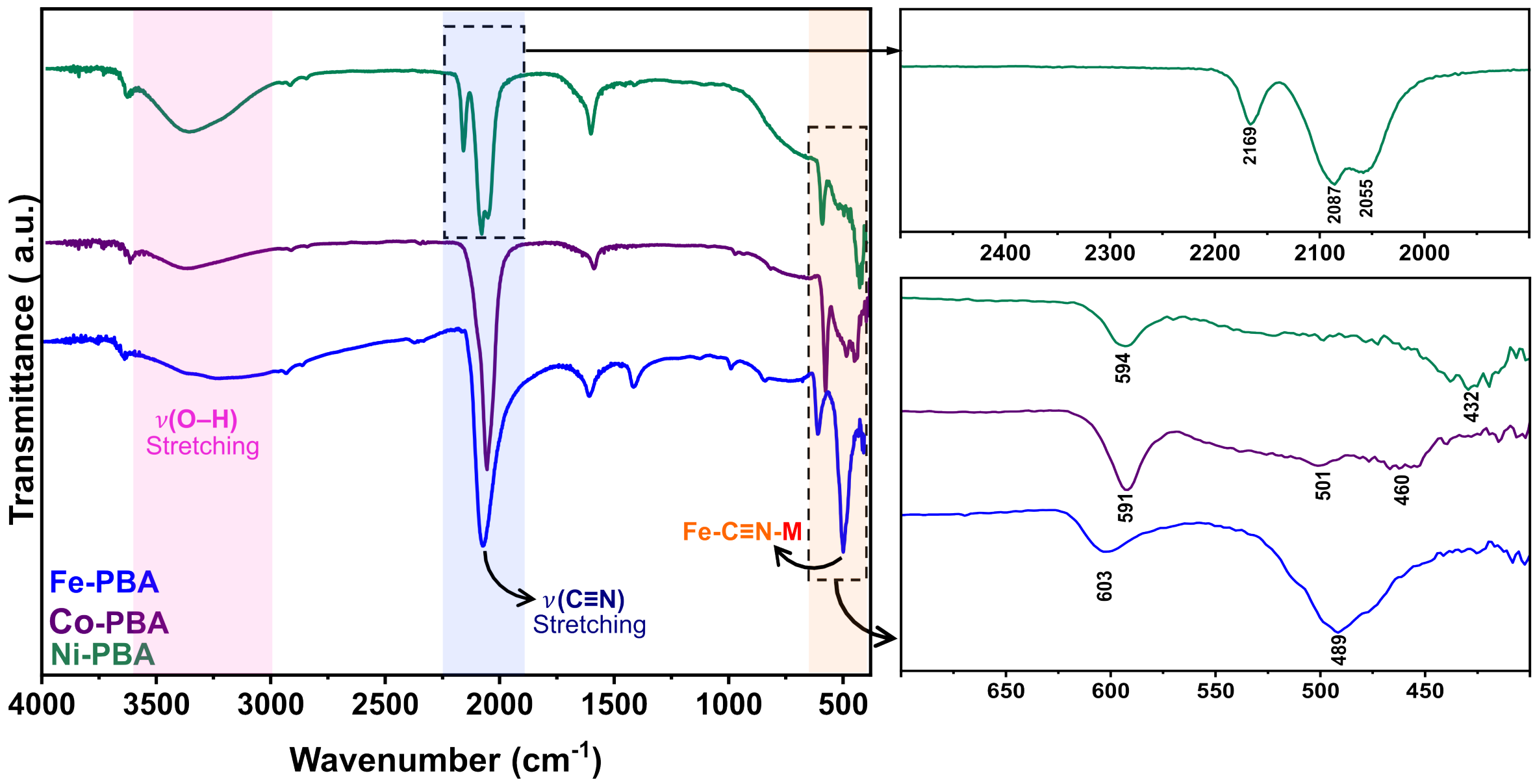

**Figure 3:** ATR-FTIR spectra of the synthesized Fe, Co, and Ni Prussian Blue analogue powders. The spectra exhibit the characteristic O–H stretching, H–O–H bending, C≡N stretching, and low-wavenumber vibrational features associated with the cyanide-bridged PBA framework.

**Table 2**
Characteristic FTIR absorption bands of the synthesized Fe-, Co-, and Ni-based Prussian Blue analogue (PBA) powders.

| Vibrational assignment | Wavenumber ($cm^{-1}$) | | |
|---|---|---|---|
| | Fe-PBA | Co-PBA | Ni-PBA |
| $\nu$(O–H), stretching | 3634, 3206 | 3630, 3381 | 3646, 3370 |
| $\delta$(H–O–H), bending | 1599 | 1604 | 1608 |
| $\nu$(C≡N), stretching | 2062 | 2071 | 2169, 2087, 2055 |
| Metal–cyanide lattice modes | 603, 489 | 591, 501, 460 | 594, 432 |

### 3.3. Raman Spectroscopy

Figure 4 presents the Raman spectra of the Fe-, Co-, and Ni-containing PBA powders deposited from the 80:20 vol% water/ethanol dispersions described in the experimental section. The principal bands are summarized in Table 3. All three compositions exhibit characteristic C≡N stretching features in the approximately 2050–2160 $cm^{-1}$ region, consistent with the cyanide-bridged PBA framework across the Fe–Co–Ni series [31, 35, 51]. The systematic changes in band position and spectral profile further indicate that the transition-metal composition affects the local vibrational environment of the framework.

Two principal C≡N stretching components are resolved for each composition in the analyzed spectra. Fe-PBA exhibits bands at approximately 2116 and 2153 $cm^{-1}$, Co-PBA at 2096 and 2136 $cm^{-1}$, and Ni-PBA at 2057 and 2089 $cm^{-1}$. Both principal components therefore shift progressively toward lower wavenumbers from Fe-PBA to Co-PBA to Ni-PBA.

The systematic evolution of the C≡N stretching modes indicates that transition-metal substitution modifies the local environment of the cyanide bridges. Since C≡N vibrational frequencies are sensitive to the metal center and its local coordination environment, as well as to structural distortions within the framework, the observed shifts are interpreted as evidence of composition-dependent changes in local structure rather than as direct indicators of a specific oxidation state or defect concentration [31, 35, 51].

Differences in spectral profile are also observed across the series. Ni-PBA exhibits a comparatively more complex C≡N region, with additional intensity between and around the two principal bands. The spectrum contains an additional feature near 2138 $cm^{-1}$, whereas Co-PBA exhibits a weaker additional contribution near 2161 $cm^{-1}$ (Table 3). Such spectral broadening or additional contributions may reflect a distribution of local vibrational environments arising from differences in coordination, hydration, lattice distortion, or structural disorder. However, these contributions cannot be separated quantitatively from Raman spectroscopy alone. The differences are therefore interpreted conservatively as variations in local vibrational heterogeneity across the series.

The Raman results complement the ATR-FTIR measurements. In the FTIR spectra, the dominant C≡N absorption shifts toward higher wavenumbers from Fe-PBA to Ni-PBA, whereas the principal Raman components shift toward lower wavenumbers. This difference does not necessarily imply conflicting structural trends. For an octahedral $Fe(CN)_6$ unit, the six C≡N stretches split into Raman-active $A_{1g}$ and $E_g$ modes and an infrared-active $T_{1u}$ mode, so the two techniques do not probe the same vibrations [31]. In addition, the Raman spectra come from micrometre-sized spots on the drop-cast deposits, whereas ATR-FTIR averages over the powder, and Prussian Blue is photosensitive under Raman excitation [35]. Selection rules do not explain the Ni-PBA spectrum, however: for a single species the Raman bands lie at or above the infrared band [31], so the 2057 $cm^{-1}$ Raman band of Ni-PBA probably comes from the additional phase or from a laser-induced change. Considered together, the two spectroscopic techniques are consistent with the view that transition-metal substitution modifies the local vibrational environment while preserving the characteristic cyanide-bridged PBA framework.

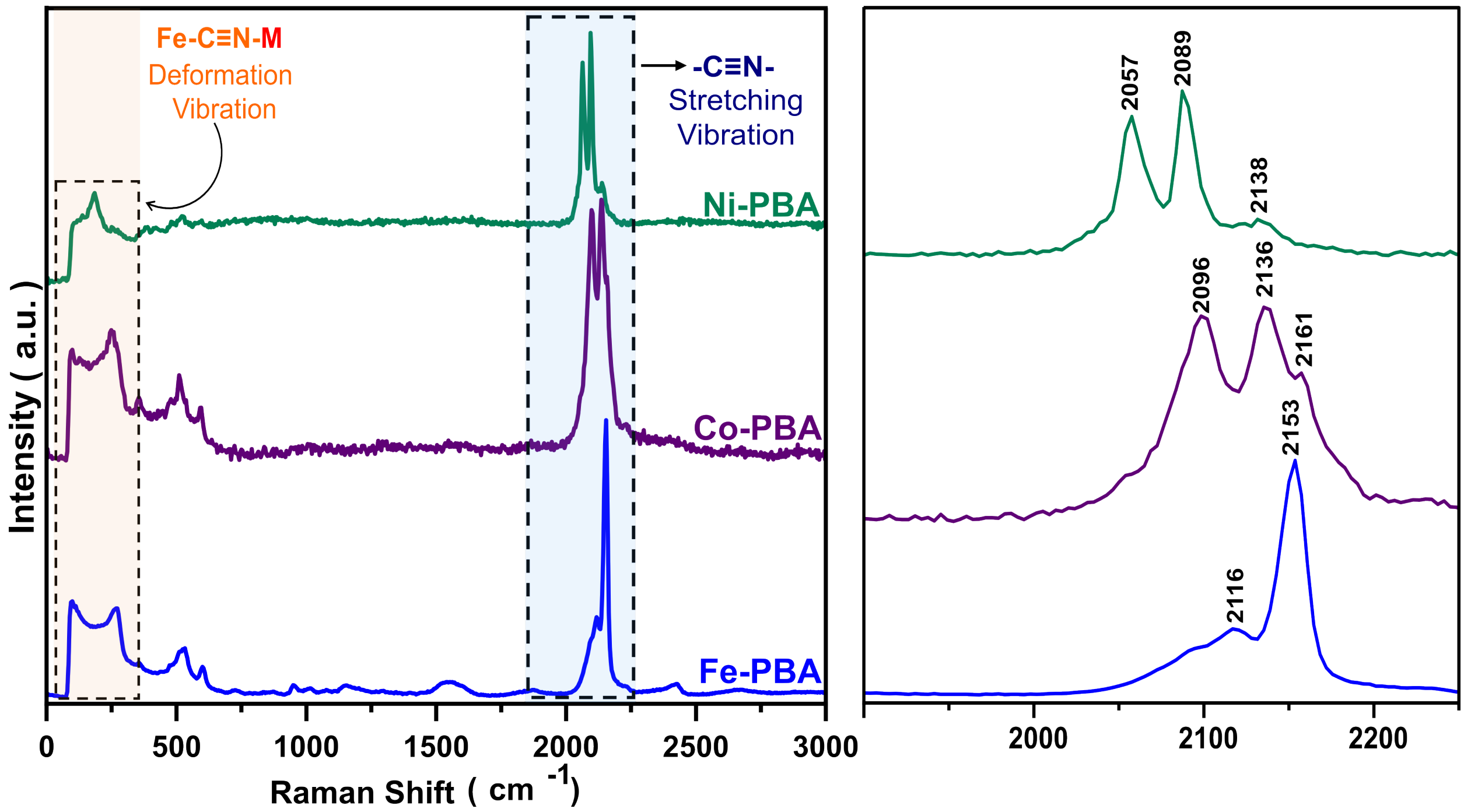

**Figure 4:** Raman spectra (532 nm excitation) of the Fe-, Co- and Ni-PBA deposits from 0 to 3000 cm$^{-1}$ (left) and enlarged C≡N stretching region (right).

**Table 3**
Characteristic Raman bands (cm$^{-1}$) observed for the synthesized Fe-, Co-, and Ni-based Prussian Blue analogue (PBA) powders.

| PBA composition | Principal band 1 | Principal band 2 | Additional band |
|---|---|---|---|
| Fe-PBA | 2116 | 2153 | – |
| Co-PBA | 2096 | 2136 | 2161 |
| Ni-PBA | 2057 | 2089 | 2138 |

## 3.4. X-ray Diffraction

Figure 5 presents the X-ray diffraction (XRD) patterns of the synthesized Fe-, Co-, and Ni-containing Prussian Blue analogue (PBA) powders, with the principal diffraction parameters summarized in Table 4. All three compositions exhibit the characteristic reflection sequence of the average cubic PBA framework, with the main reflections located at approximately 20–21°, 29°, and 41–42°. Weaker PBA reflections up to (620) are also observed (Fig. 5). The preservation of this reflection sequence across the Fe–Co–Ni series indicates that no change in the fundamental average cubic framework symmetry is resolved by powder XRD.

The principal PBA reflections were indexed as (200), (220), and (400). Small differences in peak position are observed among the three compositions, indicating changes in the average interplanar spacings and lattice dimensions. The mean lattice parameters (with the standard deviation over the three reflections) are $10.10 \pm 0.02$, $10.01 \pm 0.02$ and $10.10 \pm 0.03$ Å for Fe-, Co- and Ni-PBA, respectively. These values were estimated from only three principal reflections and should not be considered refined crystallographic parameters. The systematic increase of $a_{hkl}$ with diffraction angle observed for Fe-PBA (and smaller inconsistencies in the other samples) is consistent with an uncorrected zero/displacement contribution. A full-pattern Le Bail or Rietveld refinement, preferably using the KCl reflections as an independently verified internal reference, would provide a more robust comparison.

The measured lattice parameters show Fe- and Ni-PBA at similar values and Co-PBA at a smaller value. Given the lack of zero-shift correction and the independently unresolved effects of vacancies, interstitial cations, hydration, and local distortions, the present data do not support assigning the difference uniquely to transition-metal identity. Oxidation-state effects are another plausible contributor for Co-containing PBAs. In cobalt hexacyanoferrates, the cell shrinks from 10.2–10.3 Å with high-spin Co(II) to about 9.9 Å with low-spin Co(III) [52, 53], and 10.01 Å lies between these limits. The dominant infrared C≡N band of Co-PBA (2071 $cm^{-1}$) lies below the values reported for both Co(II)–N≡C–Fe(II) (about 2090 $cm^{-1}$) and Co(III)–N≡C–Fe(II) (about 2120 $cm^{-1}$) [53], so the present data do not settle the Co oxidation state.

Additional narrow reflections are observed at approximately 33.0, 47.4–47.5, 59.0, and 69.3°. For Co K$\alpha$ radiation ($\lambda = 1.78896$ Å) and cubic KCl with $a \approx 6.29$ Å, the expected positions are approximately 33.05° (200), 47.43° (220), 59.03° (222), and 69.34° (400). The close agreement, together with the detection of both K and Cl by EDS and the use of potassium ferrocyanide/chloride precursors, makes residual crystalline KCl the most plausible assignment of these sharp reflections. The KCl (111) reflection, expected near 28.5°, is nearly extinct because $K^+$ and $Cl^-$ are isoelectronic and scatter X-rays almost equally; the KCl reference pattern contains no odd-index reflections [54], so the PBA (220) peak is not affected. Residual KCl is therefore treated as a secondary phase arising from synthesis/washing rather than as a chloride-containing PBA phase. Additional washing with efficient solid–liquid separation would be expected to reduce this contribution when higher phase purity is required. Ni-PBA shows further sharp reflections, near 25.3, 26.0, 27.9, 33.9, 35.1, 35.6, 36.8, 38.0, 39.8, 49.9 and 57.4°, that belong neither to the PBA framework nor to KCl and indicate at least one additional crystalline phase.

Because the KCl reflections occur at well-defined positions, they may also provide an approximate internal reference for assessing sample displacement or zero-shift effects, provided that the KCl assignment is independently confirmed and the peaks are sufficiently resolved. They should not, however, be treated as a formal internal standard without such verification.

Differences in the widths and profiles of the principal PBA reflections are also observed across the series. Co-PBA exhibits comparatively narrower reflections, whereas Ni-PBA shows more pronounced broadening of the (400) reflection. Such variations may reflect differences in coherent diffraction-domain size, microstrain, structural disorder, hydration, or other local structural effects. Because instrumental broadening was not independently determined and no size–strain analysis was performed, the FWHM values are used only for qualitative comparison and are not interpreted as direct measures of crystallite size or defect concentration.

The XRD patterns confirm a common average cubic PBA framework for the Fe-, Co-, and Ni-containing powders, while revealing composition-dependent variations in lattice dimensions and diffraction profiles. These differences provide an average-structure context for the local vibrational features observed by FTIR and Raman spectroscopy and for the thermal evolution discussed below. The XRD measurements characterize the as-synthesized powders and therefore establish the initial crystalline state rather than the phases that may develop during subsequent thermal treatment.

**Table 4**
Principal X-ray diffraction (XRD) reflections of the synthesized Fe-, Co-, and Ni-based Prussian Blue analogue (PBA) powders. The interplanar spacings and lattice parameters were calculated using Co K$\alpha$ radiation ($\lambda = 1.78896$ Å).

| PBA | $(hkl)$ | $2\theta$ (°) | FWHM (°) | $d_{hkl}$ (Å) | $a_{hkl}$ (Å) |
|---|---|---|---|---|---|
| Fe-PBA | (200) | 20.438 | 0.793 | 5.0418 | 10.0837 |
| | (220) | 29.038 | 1.016 | 3.5679 | 10.0916 |
| | (400) | 41.388 | 0.851 | 2.5312 | 10.1249 |
| Co-PBA | (200) | 20.638 | 0.488 | 4.9935 | 9.9870 |
| | (220) | 29.238 | 0.651 | 3.5440 | 10.0240 |
| | (400) | 41.838 | 0.606 | 2.5052 | 10.0208 |
| Ni-PBA | (200) | 20.488 | 0.606 | 5.0297 | 10.0594 |
| | (220) | 28.938 | 0.775 | 3.5800 | 10.1257 |
| | (400) | 41.488 | 0.938 | 2.5254 | 10.1016 |
| | | | **Mean lattice parameter, $\bar{a}$ (Å)**, ± standard deviation | | |
| | | | | Fe-PBA | 10.10 ± 0.02 |
| | | | | Co-PBA | 10.01 ± 0.02 |
| | | | | Ni-PBA | 10.10 ± 0.03 |

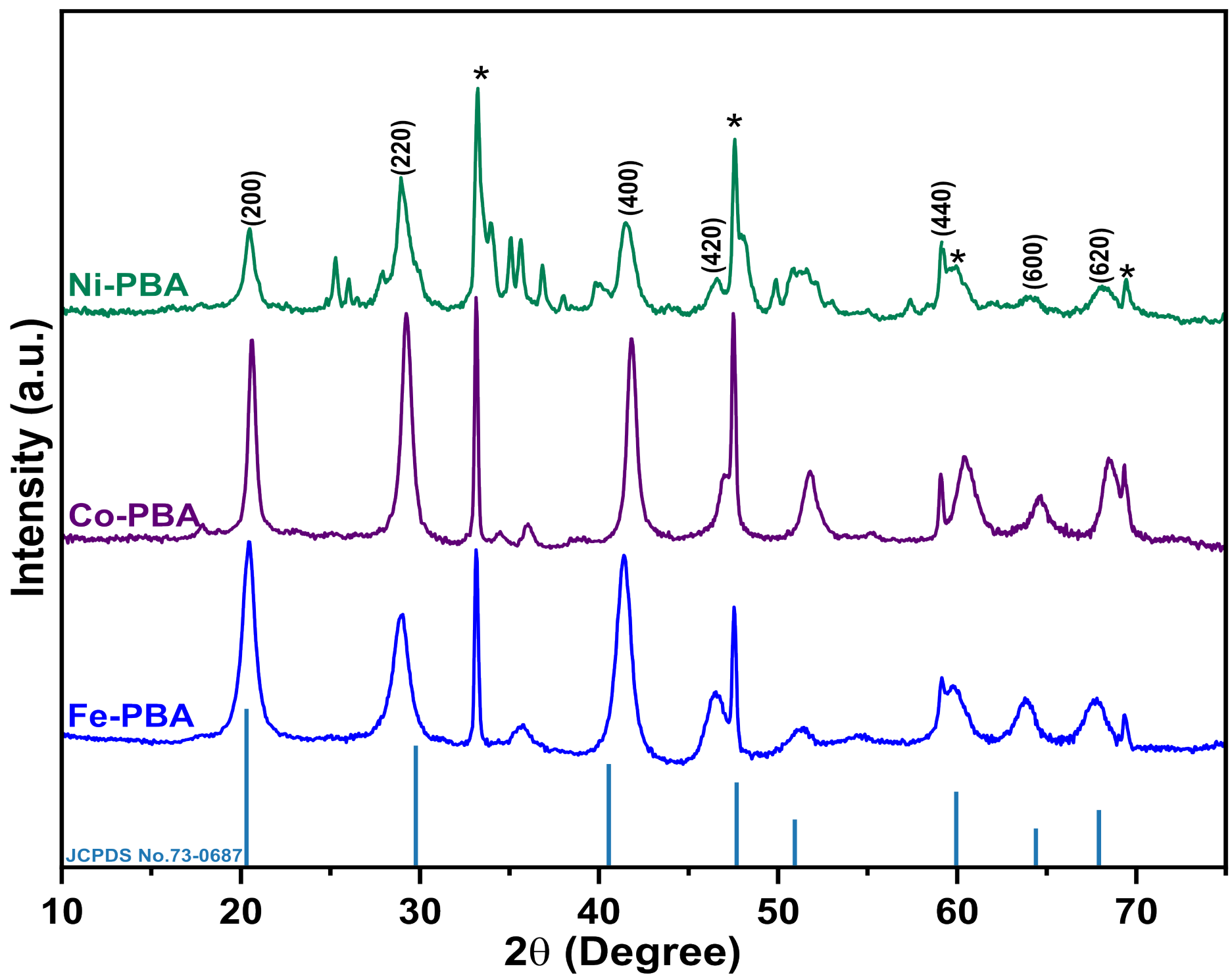

**Figure 5:** XRD patterns of the Fe-, Co-, and Ni-containing Prussian Blue analogue powders recorded using Co K$\alpha$ radiation ($\lambda = 1.78896$ Å). The principal reflections of the cubic PBA framework are indexed according to their $(hkl)$ planes. Asterisks mark reflections of residual KCl.

### 3.5. Thermal Mass-Loss Behavior and Thermal Evolution

The thermal behavior of Fe-, Co-, and Ni-containing Prussian Blue analogue (PBA) powders was investigated by thermogravimetric analysis (TG) and single differential thermal analysis (SDTA) under flowing nitrogen. The corresponding TG curves and SDTA signals are shown in Fig. 6.

All three compositions exhibit progressive mass loss with increasing temperature, although the magnitude and distribution of the thermal features differ substantially. Fe-PBA shows the most pronounced mass decrease at low temperature, whereas Co-PBA and Ni-PBA exhibit comparatively small initial changes. Because the individual features overlap over composition-dependent temperature ranges, they are considered here as selected mass-loss regions rather than as discrete decomposition steps.

For the selected low-temperature regions, Fe-PBA decreases from approximately 99.57% to 90.03%, corresponding to a mass loss of about 9.6%. The corresponding losses for Co-PBA and Ni-PBA are approximately 0.8% and 0.6%, respectively. Additional mass-loss regions of approximately 7.5% are observed for Fe-PBA, 2.3% for Co-PBA, and 2.9 and 2.5% for Ni-PBA. These values are summarized in Table 5. Since the selected regions do not cover identical temperature intervals for all three compositions, the percentages should be regarded as feature-specific mass losses rather than directly comparable measures of total hydration. Nevertheless, the substantially larger initial loss of Fe-PBA indicates a stronger low-temperature volatile-loss response under the present measurement conditions.

The low-temperature mass loss is consistent with the release of water-associated species. Water may occur as surface-adsorbed, interstitial, or coordination-associated species in PBA structures, and their release can overlap over a broad temperature range. The presence of water in the prepared powders is independently supported by the O–H stretching and H–O–H bending bands observed by FTIR. Similar multistage dehydration behavior has been reported for Prussian Blue and related PBAs under inert and oxidative conditions [55, 56]. Chen et al. reported water-related mass losses extending to approximately 300 °C for FeHCF, CoHCF, NiHCF, CuHCF, and ZnHCF under $N_2$, with contributions attributed to physically adsorbed, interstitial, and coordinated water [57]. The present TG profiles are consistent with such a multicomponent dehydration process; however, the individual mass-loss regions cannot be quantitatively assigned to specific water populations from TG data alone.

The magnitude of the observed losses falls within the broader range reported for PBA materials, although direct numerical comparison requires consideration of composition, synthesis conditions, atmosphere, and temperature interval. For example, Chen et al. reported total water losses of 27.6, 30.7, and 32.0 wt% for FeHCF, CoHCF, and NiHCF, respectively, under $N_2$ [57]. Li et al. reported losses of 21.17, 29.43, and 9.94% over the first two TG steps of NiFe-, CoFe-, and MnFe-PBA, respectively, with total losses of 46.47, 41.10, and 30.37% up to 900 °C [49]. The selected low-temperature losses measured in the present work are therefore not interpreted as total water contents. Rather, they demonstrate that both hydration and its thermal release depend strongly on PBA composition and preparation history.

Above approximately 200 °C, all three powders continue to lose mass, indicating that the thermal evolution extends beyond dehydration. Fe-PBA exhibits the largest cumulative decrease, whereas Ni-PBA retains the greatest fraction of its initial mass. Continued mass loss at elevated temperatures has also been reported for Prussian Blue-type materials and mixed-metal PBAs under inert conditions, where dehydration is followed by further structural evolution and decomposition of the cyanide-containing framework [55, 56]. The present measurements therefore indicate progressive high-temperature transformation, although the individual processes cannot be uniquely assigned from TG/SDTA alone.

At 600 °C, the retained masses are approximately 65.7%, 70.3%, and 77.1% for Fe-PBA, Co-PBA, and Ni-PBA, respectively, corresponding to total mass losses of 34.3, 29.7, and 22.9%. These values describe the measured samples but should not be interpreted as a direct ranking of intrinsic thermal stability. The final retained mass can be affected by residual KCl or other non-volatile inorganic phases, as well as by the initial volatile content, framework composition, defect population, local coordination environment, and thermally generated products. Consequently, differences in final residue cannot be attributed to a single structural parameter.

The simultaneously recorded SDTA signals provide complementary information on the thermal events. The most pronounced variations occur in the low-temperature region, coinciding with the region where the TG curves exhibit the strongest volatile mass loss. Fe-PBA shows the largest response, whereas Co-PBA and Ni-PBA exhibit smaller, composition-dependent features. Two features—a step near 70 °C and a maximum near 175 °C—occur at approximately the same temperatures for all three samples. Their reproducibility across compositions suggests a possible instrumental contribution rather than a composition-specific thermal event. A blank measurement using an empty crucible would therefore be useful to determine their origin.

The thermal behavior is broadly consistent with the differences observed by vibrational spectroscopy. FTIR confirms the presence of molecular water, while the systematic changes in the C≡N stretching region observed by FTIR and Raman indicate composition-dependent modifications of the local cyanide environment. These structural differences may contribute to the distinct hydration and thermal responses observed for the three compositions, although the present data do not establish a direct quantitative relationship between an individual vibrational feature and a specific mass-loss process. In particular, the larger low-temperature loss observed for Fe-PBA is consistent with the structural characteristics of potassium-poor Prussian Blue, $Fe_4[Fe(CN)_6]_3$, in which approximately one quarter of the $[Fe(CN)_6]$ sites are vacant and the resulting exposed Fe(III) sites can bind water [2]. The Co- and Ni-containing analogues, prepared from M(II) precursors, are expected to have different vacancy and coordination characteristics, which may contribute to their distinct thermal profiles.

Finally, the present thermal analysis has several limitations. No replicate measurements were performed, and the initial sample masses differed (12.48, 9.75, and 6.91 mg), which may influence heat transfer and the apparent shape of thermal features. In addition, TG/SDTA does not directly identify the chemical nature of the evolved species. Evolved-gas analysis, such as TG–MS, would therefore be required to distinguish water release from the evolution of cyanide-containing or other volatile species. Accordingly, the present results are interpreted as evidence for composition-dependent dehydration and subsequent thermal evolution rather than as a complete assignment of the individual decomposition mechanisms.

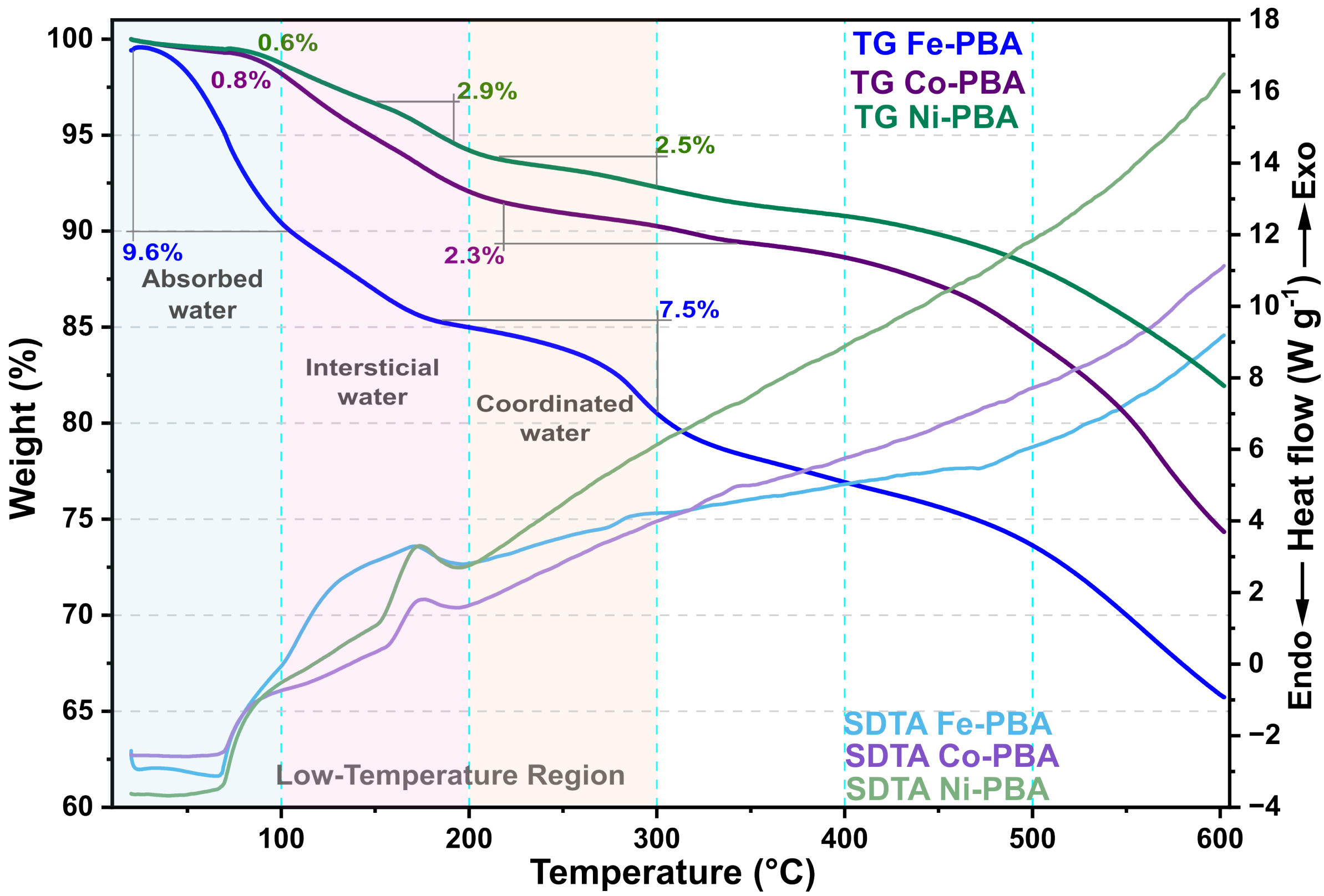

**Figure 6:** TG curves and SDTA signals of the Fe-, Co-, and Ni-containing Prussian Blue analogue powders measured under flowing $N_2$ from 25 to 600 °C at a heating rate of 10 °C $min^{-1}$.

**Table 5**
Thermogravimetric characteristics of the Fe-, Co-, and Ni-PBA powders. The selected and additional losses correspond to the mass-loss features indicated in Fig. 6; they are not intended as decomposition stoichiometry. Mass retained and total loss at 600 °C are relative to the initial sample mass.

| Sample | Initial mass (mg) | Selected initial loss (%) | Additional regions (%) | Mass retained at 600 °C (%) | Total loss at 600 °C (%) |
|---|---|---|---|---|---|
| Fe-PBA | 12.48 | 9.6 | 7.5 | 65.7 | 34.3 |
| Co-PBA | 9.75 | 0.8 | 2.3 | 70.3 | 29.7 |
| Ni-PBA | 6.91 | 0.6 | 2.9, 2.5 | 77.1 | 22.9 |

**Table 6**
Comparison of thermogravimetric characteristics for selected Prussian Blue analogue materials reported in the literature and in the present work. Because the temperature intervals, atmospheres, compositions, and synthesis conditions differ, the reported values should not be interpreted as directly equivalent quantities.

| Study / material | Atmosphere | Mass-loss region | Reported loss | Higher-temperature behavior |
|---|---|---|---|---|
| This work — Fe-PBA | $N_2$ | selected low-temperature feature | 9.6%[a] | 34.3% total mass loss at 600 °C |
| This work — Co-PBA | $N_2$ | selected low-temperature feature | 0.8%[a] | 29.7% total mass loss at 600 °C |
| This work — Ni-PBA | $N_2$ | selected low-temperature feature | 0.6%[a] | 22.9% total mass loss at 600 °C |
| Chen et al. — FeHCF | $N_2$ | < 300 °C | 27.6%[b] | Further thermal evolution |
| Chen et al. — CoHCF | $N_2$ | < 300 °C | 30.7%[b] | Further thermal evolution |
| Chen et al. — NiHCF | $N_2$ | < 300 °C | 32.0%[b] | Further thermal evolution |
| Chen et al. — CuHCF | $N_2$ | < 300 °C | 37.0%[b] | Further thermal evolution |
| Chen et al. — ZnHCF | $N_2$ | < 300 °C | 15.4%[b] | Relatively high mass retention |
| Li et al. — NiFe-PBA | Air | < 350 °C | 21.17%[c] | Marked mass loss above 350 °C |
| Li et al. — CoFe-PBA | Air | < 350 °C | 29.43%[c] | Marked mass loss above 350 °C |
| Li et al. — MnFe-PBA | Air | < 350 °C | 9.94%[c] | Marked mass loss above 350 °C |
| Domonov et al. — Ni and Cu hexacyanoferrates and hexacyanocobaltates | Ar | Low-temperature region | Up to ~33%[d] | Mass loss continues at higher temperatures |

[a] Initial-region mass loss; not total water content.
[b] Total water loss reported by Chen et al. [57].
[c] Sum of the first two TG steps reported by Li et al. [49]; the total losses are 46.47, 41.10 and 30.37%.
[d] Water-related mass loss up to ~33% reported by Domonov et al. [56].

## 4. Conclusion

Fe-, Co-, and Ni-containing Prussian Blue analogue powders were synthesized by aqueous co-precipitation under identical conditions and characterized by FEG-SEM/EDS, ATR-FTIR, Raman spectroscopy, XRD, and TG/SDTA. The results support formation of cyanide-bridged PBA-type frameworks across all three compositions while revealing differences in aggregate morphology, local vibrational environment, apparent lattice dimensions, and thermal evolution. The data do not by themselves establish the complete bulk chemical formulas or oxidation-state distributions.

The additive-free synthesis route enables a direct comparison of the three compositions without complexing agents, surfactants, or capping species. The powders exhibit pronounced aggregation, while the interconnected fissures observed after drop casting are attributed primarily to solvent evaporation and drying-induced stresses within the particulate deposits. K and Cl detected by EDS, together with the sharp XRD reflections, show residual KCl in all three powders. The lower potassium content and larger low-temperature mass loss of Fe-PBA are consistent with its synthesis from Fe(III); Co-PBA, made from Co(II), retains more potassium. Ni-PBA also contains an unidentified crystalline phase that may hold part of its potassium.

XRD shows a common average cubic PBA-type framework, with apparent lattice parameters of $10.10 \pm 0.02$, $10.01 \pm 0.02$ and $10.10 \pm 0.03$ Å for Fe-, Co- and Ni-PBA, before correction for the zero offset. FTIR and Raman spectroscopy reveal systematic but technique-dependent changes in the C≡N stretching region, indicating composition-dependent modifications of the local cyanide environment. TG/SDTA further shows distinct low-temperature mass-loss behavior and different mass retention at 600 °C under $N_2$. These thermal differences depend on the initial volatile content, framework composition, and products formed during heating and therefore are not interpreted as an intrinsic stability ranking.

The present study provides a comparative dataset for Fe-, Co-, and Ni-containing PBA powders prepared by the same additive-free route. ICP-OES and CHN analysis are needed to establish the formulas and a full search-match of the diffraction data to identify the additional phase in Ni-PBA.

## Declaration of competing interest

The authors declare that they have no known competing financial interests or personal relationships that could have appeared to influence the work reported in this paper.

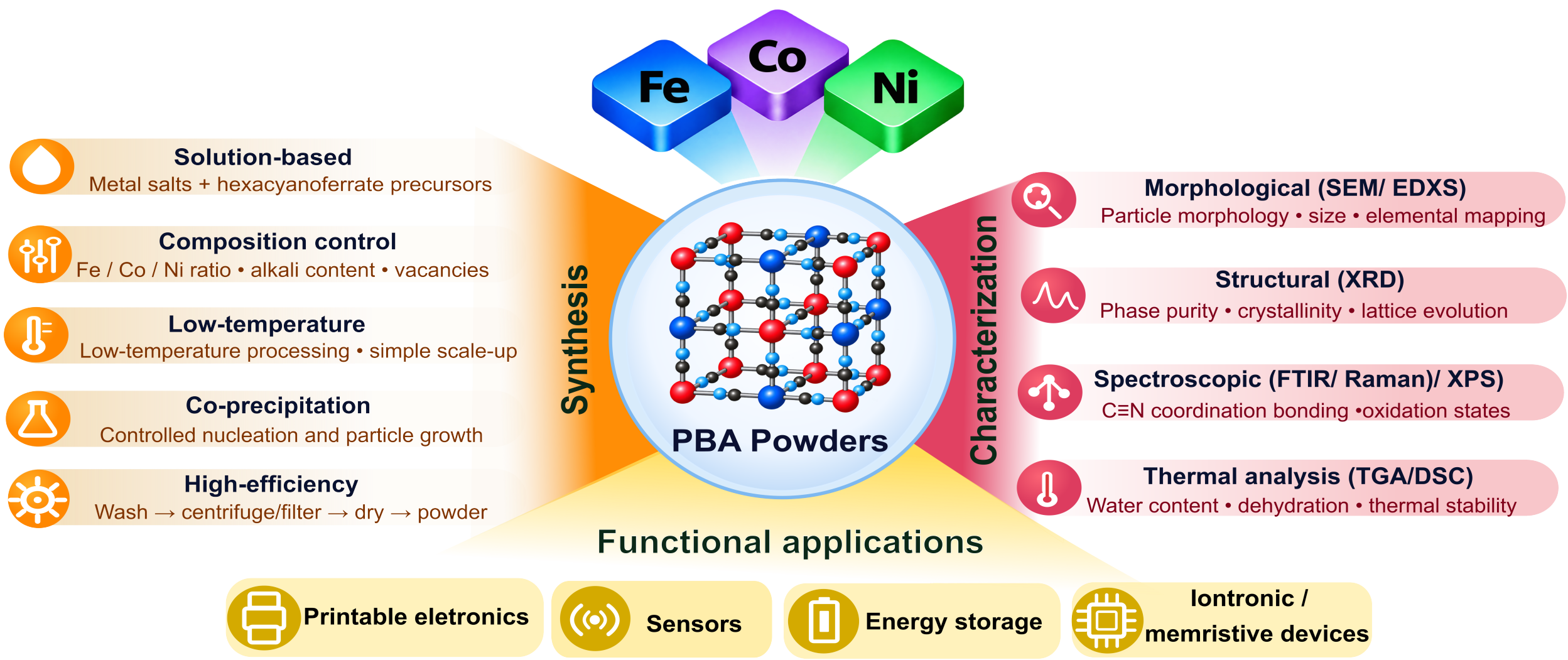
Fe
Co
Ni
Solution-based
Metal salts + hexacyanoferrate precursors
Composition control
Fe / Co / Ni ratio • alkali content • vacancies
Low-temperature
Low-temperature processing • simple scale-up
Co-precipitation
Controlled nucleation and particle growth
High-efficiency
Wash → centrifuge/filter → dry → powder
Synthesis
PBA Powders
Characterization
Morphological (SEM/ EDXS)
Particle morphology • size • elemental mapping
Structural (XRD)
Phase purity • crystallinity • lattice evolution
Spectroscopic (FTIR/ Raman)/ XPS)
C≡N coordination bonding •oxidation states
Thermal analysis (TGA/DSC)
Water content • dehydration • thermal stability
Functional applications
Printable eletronics
Sensors
Energy storage
Iontronic /
memristive devices